\documentclass[twocolumn,floatfix,superscriptaddress,a4paper,showkeys,nofootinbib,reprint]{revtex4-1}

\usepackage[colorlinks=true,linktocpage=true,linkcolor=blue,citecolor=blue,allcolors=blue]{hyperref}
\usepackage{epsfig}
\usepackage{latexsym}
\usepackage[utf8]{inputenc}
\usepackage{xspace}
\usepackage{indentfirst}
\usepackage{enumitem}
\usepackage{color}

\usepackage{setspace}
\usepackage{lipsum}

\usepackage{todonotes}

\usepackage{amsmath}
\usepackage{amssymb}
\usepackage[english]{babel}
\usepackage{url}

\newcommand{\mean}[1]{\langle #1 \rangle}

\newcommand\ddfrac[2]{\frac{\displaystyle #1}{\displaystyle #2}}

\newcommand{\eq}[1]{\begin{align} #1 \end{align}}

\begin{document}

\title{
Bose-Einstein condensation phenomenology in systems with repulsive interactions
}
\author{Oleh Savchuk}
\affiliation{Physics Department, Taras Shevchenko National University of Kyiv, 03022 Kyiv, Ukraine}
\author{Yehor Bondar}
\affiliation{Physics Department, Taras Shevchenko National University of Kyiv, 03022 Kyiv, Ukraine}
\author{Oleksandr Stashko}
\affiliation{Physics Department, Taras Shevchenko National University of Kyiv, 03022 Kyiv, Ukraine}
\author{Roman V. Poberezhnyuk}
\affiliation{Bogolyubov Institute for Theoretical Physics, 03680 Kyiv, Ukraine}
\affiliation{Frankfurt Institute for Advanced Studies, Giersch Science Center, D-60438 Frankfurt am Main, Germany}
\author{Volodymyr Vovchenko}
\affiliation{Nuclear Science Division, Lawrence Berkeley National Laboratory, Berkeley, CA 94720, USA}
\affiliation{Frankfurt Institute for Advanced Studies, Giersch Science Center, D-60438 Frankfurt am Main, Germany}
\author{Mark I. Gorenstein}
\affiliation{Bogolyubov Institute for Theoretical Physics, 03680 Kyiv, Ukraine}
\affiliation{Frankfurt Institute for Advanced Studies, Giersch Science Center, D-60438 Frankfurt am Main, Germany}

\author{Horst Stoecker}
\affiliation{Frankfurt Institute for Advanced Studies, Giersch Science Center,
D-60438 Frankfurt am Main, Germany}
\affiliation{Institut f\"ur Theoretische Physik,
Goethe Universit\"at Frankfurt, D-60438 Frankfurt am Main, Germany}
\affiliation{GSI Helmholtzzentrum f\"ur Schwerionenforschung GmbH, D-64291 Darmstadt, Germany}

\date{\today}

\begin{abstract}
The role of repulsive interactions 
in  statistical systems of Bose particles is investigated.
Three different phenomenological frameworks are considered: 
a mean field model, an excluded volume model, and a model with a medium  dependent effective mass.
All three models are tuned to yield similar equations of state, with only minor deviations from the ideal Bose gas at small chemical potentials. 
Our analysis indicates, however, that 
these models lead to qualitatively different results for the Bose-Einstein condensation phenomenon. 
We discuss the different aspects of this phenomenon, namely, 
an onset of the Bose-Einstein condensation, 
particle number fluctuations, 
and a behavior of the Bose condensate.  
The obtained results can be helpful for interpreting the lattice QCD data at small temperature and large isospin chemical potential and the data on multiple pion production in high energy nuclear collisions.

\end{abstract}
\keywords{Bose-Einstein condensation, repulsive interactions, equation of state}

\maketitle

\section{Introduction}
The Bose-Einstein condensation (BEC) in the ideal gas of bosons was predicted many years ago~\cite{Bose:1924mk,einstein1925stizunger}
and later experimentally confirmed for cold atomic gases in magnetic traps
\cite{Anderson198,PhysRevLett.75.1687,PhysRevLett.75.3969,RevModPhys.71.463}.
This effect appears to be common for different systems of free or interacting bosons, ranging from condensed matter physics to high-energy nuclear physics and astrophysics~(see, e.g., Refs.~\cite{Satarov:2017jtu,Begun:2006gj,Begun:2008hq,Strinati_2018,Nozieres:1985zz,PhysRevLett.101.082502,Chavanis:2011cz,Mishustin_2019}).

The theory of the BEC phenomenon for interacting particles has been extensively discussed \cite{kapusta_gale_2006,Andersen_2004,griffin1996bose,PhysRevA.88.053633,Watabe_2019}.
In particular, modifications of the BEC onset line~(further referred to as the BEC-line) due to the small repulsive interactions between particles were predicted~\cite{PhysRevLett.83.1703,Baym_2000, Holzmann_1999,Holzmann_2001,Baym_1999,PhysRevLett.83.3770,PhysRev.91.1291,PhysRev.91.1301}.
Namely,
an increase of the temperature of the onset of BEC
due to the repulsive interactions 
when compared with ideal Bose gas (Id-BG) at the same density was found.
 However, this 
conclusion is not of general validity, as will be shown in the present paper.  
We use three different phenomenological models to describe the effects of particle repulsion in boson systems: a mean field model, an excluded volume model, and a model with a medium dependent effective mass. 
The parameters of these models
are tuned to produce quantitatively similar equations of state at zero chemical potential and with only small deviations from the Id-BG. However, the properties of the BEC phenomenon 
appear to be very sensitive to specific features of these considered models. 
This fact motivates our interest to perform a comparative analysis of these models. 

Most real systems have, in addition to repulsion, also attractive forces that dominate thermodynamics at low densities, producing phase diagrams with a more complex structure compared to the ones obtained here. 
In particular, one can observe 1st and 2nd order phase transitions in addition to the BEC. 
In the phase diagram regions where such effects can be neglected, however, we expect our arguments to be true.

In addition to the behavior of 
the BEC-line and  the phase with the Bose condensate (BC), 
we also analyze the behavior of particle number fluctuations, as their measurements can serve as a signature of the BEC.
We find qualitatively different results for all considered quantities for the BEC in the three considered models.

Our discussion is appropriate for a generic system of bosons with repulsive interactions. Nevertheless, 
to be specific 
we will refer mostly to a statistical system of $\pi$-mesons. 
Two arguments motivate this choice. 
First, recent results from lattice QCD support an existence of the pion BEC at finite isospin chemical potential
~\cite{Brandt:2017oyy,Brandt_2018}, as suggested earlier by the chiral perturbation theory~\cite{Son:2000xc}.
Second, the pion BEC phenomenon has a number of applications, including ultra-relativistic collisions of heavy ions~\cite{Begun:2006gj,Begun:2008hq}, the hypothetical pion stars
\cite{Brandt:2018bwq,Mannarelli_2019,andersen2018boseeinstein}, and the cosmic trajectory in the early universe~\cite{Abuki:2009hx,Brandt:2017oyy}. Recently, the possibility of Bose condensation in a pion system was
considered within a Skyrme-like model including both attractive and repulsive interaction terms \cite{Anchishkin_2019}.

The paper is organized as follows. Section~\ref{sec-ideal} describes the characteristics of the BEC in the Id-BG gas. 
Section~\ref{sec-frameworks} presents a description of the theoretical frameworks used in the paper. In Sec.~\ref{sec-results} we present  the model results for the BEC and reveal the qualitative differences obtained within the three considered models. A summary in Sec.~\ref{sec-concl} closes the paper.

\section{Ideal Bose Gas}
\label{sec-ideal}
The pressure function  of the relativistic gas in the grand canonical ensemble can be written as  \cite{GNS}
\eq{\label{p-id}
p_{\rm id}(T,\mu)=\frac{ d}{6\pi^2} \int_0^{\infty}
dk
\frac{ k^4}{\sqrt{k^{2} + m^2}}\, f_{\rm k}~,
}
where the momentum distribution $f_k$ reads
\eq{\label{fk}
f_{\rm k}
(T,\mu;m)
=\left[{\rm \exp}\left(\frac{
\sqrt{k^2+m^2}-\mu}{T}\right)-\eta\,\right]^{-1}~,
}
where $\eta=1$ and $\eta=-1$ for the Bose and Fermi statistics, respectively. The classical Boltzmann approximation corresponds to $\eta=0$.
$m$ is the particle mass,
$T$ and $\mu$ are the system's temperature and chemical potential, respectively, and $d$ is the degeneracy factor.
The density of particles in the ideal gas is given by
\eq{\label{n-id}
n_{\rm id}(T, \mu) \equiv \left(\frac{\partial p_{\rm id}}{\partial \mu}\right)_T= \frac{d}{2\pi^2}\int\limits_{0}^{\infty}dk~k^2~f_{\rm k}(T,\mu;m)~.
}
In what follows we discuss the identical bosons in the same internal state, spin and iso-spin states. Thus we keep $d=1$ for the number of internal degrees of freedom.

At  fixed $T$, the particle number density ({\ref{n-id})} is a monotonously increasing function of $\mu$. For bosons, $\eta=1$, 
the integral in Eq.~(\ref{n-id}) reaches  its maximal value at $\mu=m$. 
Chemical potentials values larger than $\mu=m$ are forbidden as they would lead to negative values of particle occupation numbers $f_k$ in some $k$-states. 
Note that such a restriction on $\mu$ is absent in the ideal Fermi gas with $\eta=-1$.  
At $\mu=m$ the total number of particles $N_0$ at $k=0$ may become of a macroscopic magnitude, i.e. proportional to the system's volume, $N_0\propto V$. 
In this case the particle number density $n_0$ in the lowest energy level, $k=0$, should be accounted separately,
as an additional term in the particle number density. 
The total particle number density $n$ is then written as follows:
\eq{n=n_{\rm id}(T,\mu=m)+n_0,}
where $n_0\ge 0$ is the density of particles with zero momentum,
the so-called BC density.

\begin{figure*}
\includegraphics[width=.49\textwidth]{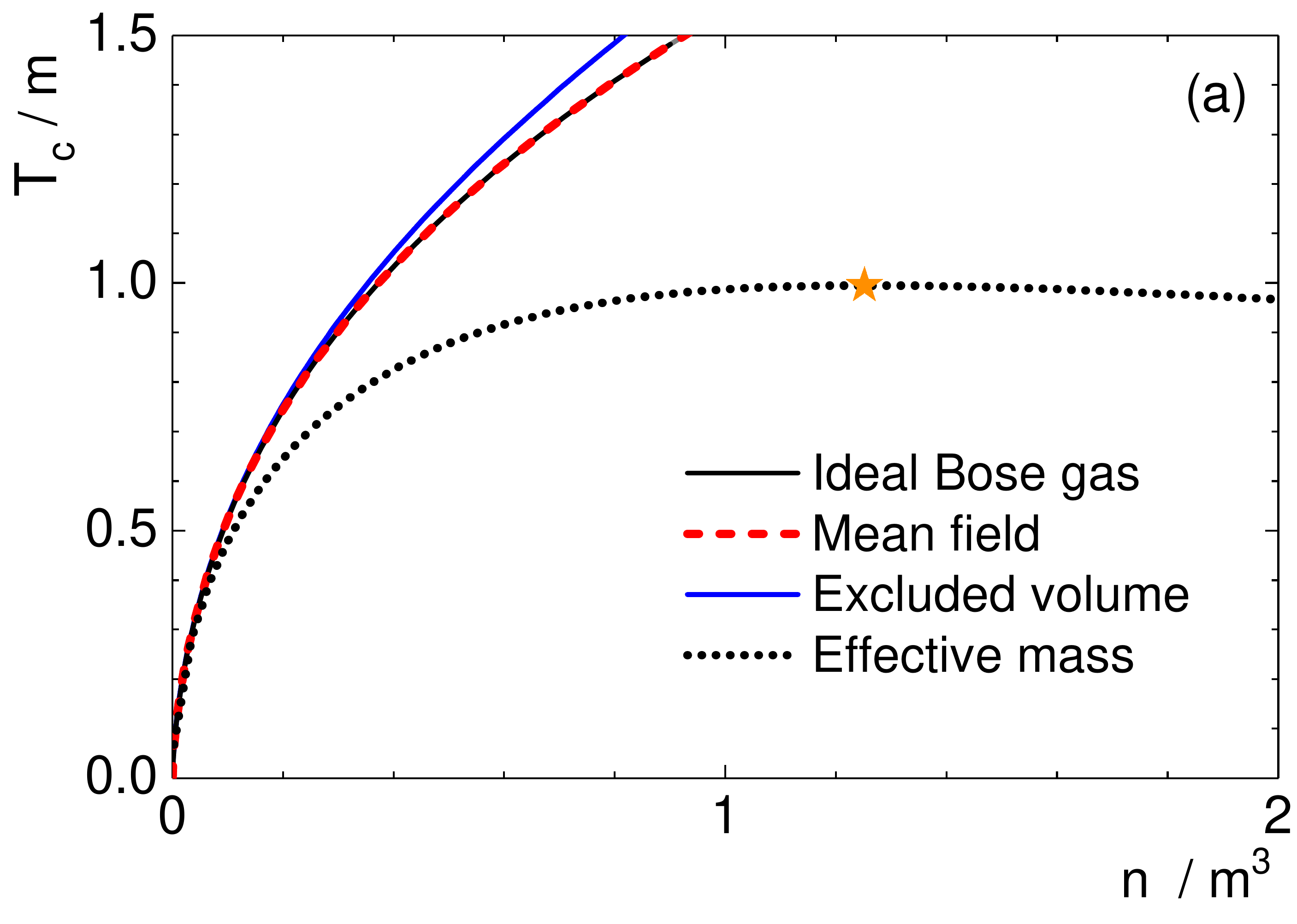}
\includegraphics[width=.49\textwidth]{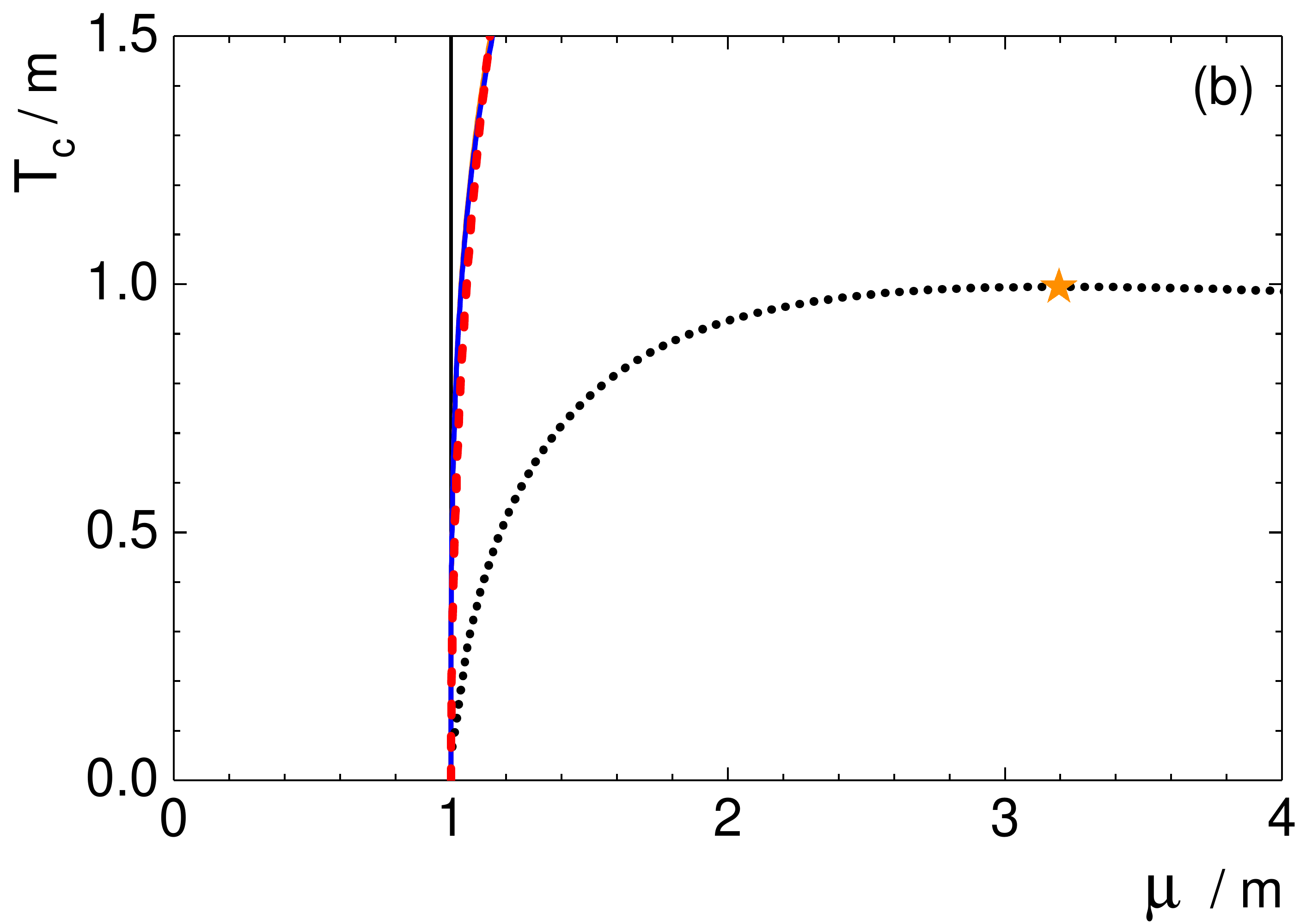}
\caption{\label{cond-lines}
The BEC-lines $T_c=T_c(n)$ (a) and $T_c=T_c(\mu)$ (b)
for Id-BG, MF, EV, and EM models.
The star denotes
the maximal temperature $T_c^{\rm max}$ at which the BEC is possible in the EM model.
}
 \end{figure*}

The BEC-line, $T_c=T_c(n)$,
can be obtained by substituting $\mu=m$ in Eq.~(\ref{n-id}) and solving the equation with respect to $T$. 
For the non-relativistic, $(k^2+m^2)^{1/2}\approx m+k^2/(2m)$, and ultra-relativistic, $(k^2+m^2)^{1/2}\approx k$, approximations the solutions are explicit~\cite{LL,GNS}:
\eq{
& T_c\approx \frac{2\pi}{m}\left(\frac{n}{\zeta(3/2   )}\right)^{\frac{2}{3}}\approx 3.31~\frac{n^{2/3}}{m},~~  T_c/m\ll 1, \label{nrel}\\
& T_c\approx \left(\frac{\pi^2 n}{\zeta(3)}\right)^{\frac{1}{3}}\approx\,
2.02~n^{1/3},~~~~~~~~~ T_c/m\gg 1~,\label{rel}
}
where $\zeta(x)=\sum_{n=1}^\infty n^{-x}$ is the Riemann zeta function with $\zeta(3/2)\approx 2.612$ and $\zeta(3)\approx 1.202$.
The BEC in the cold low-density atomic gases corresponds to region $T/m < 10^{-10}$, while in  nuclear physics, e.g. for $\alpha$-particles,  to $T/m< 10^{-3}$. Thus, these physical phenomena can be accurately described within the non-relativistic limit (\ref{nrel}). However, the BEC of pions can happen at $T/m_\pi \sim 1$ and this necessitates using the relativistic formulation. The Id-BG BEC-line $T_c(n)$ is shown by the solid line in Fig.~\ref{cond-lines} (a).   

\begin{figure*}
\includegraphics[width=.49\textwidth]{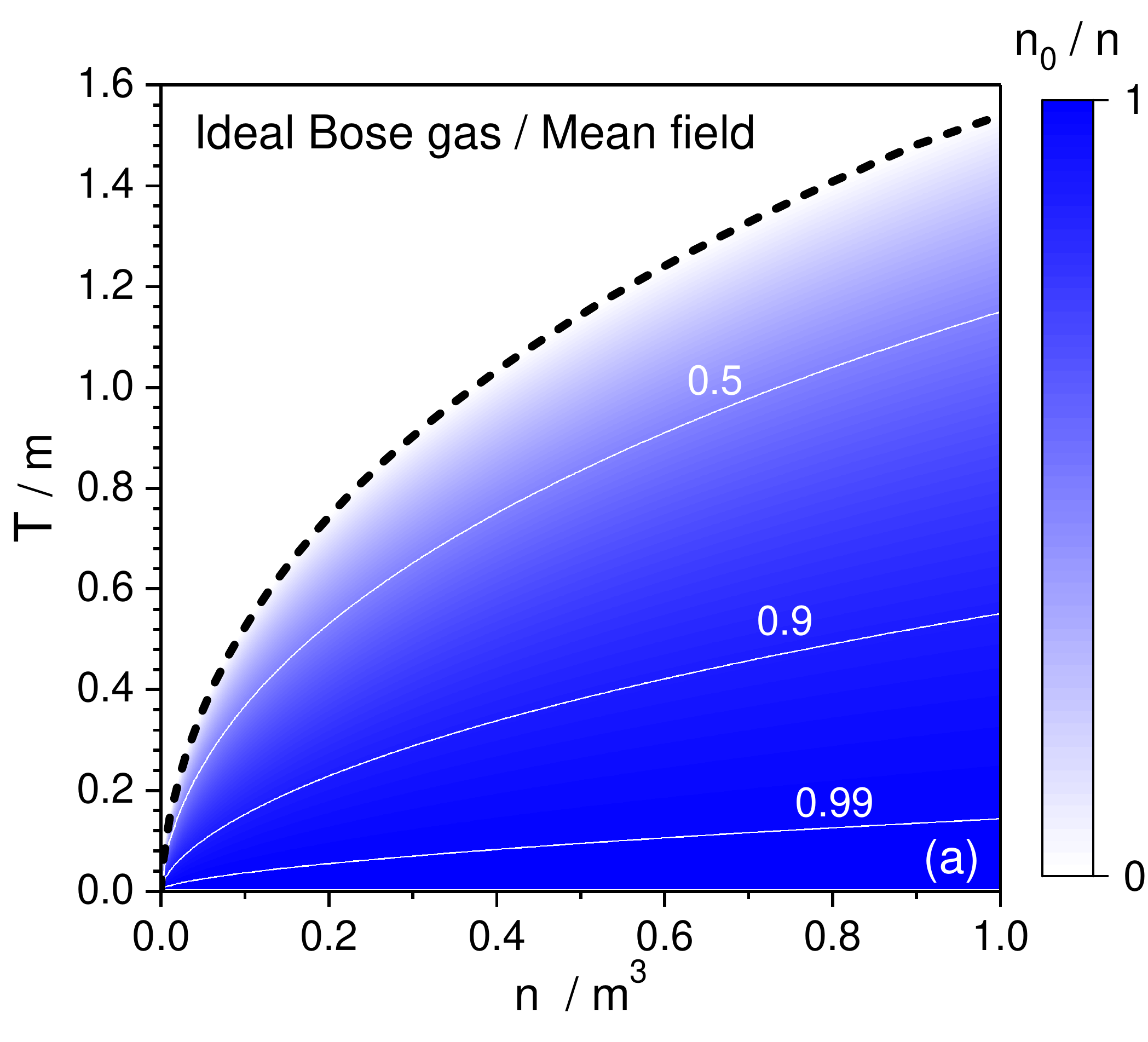}
\includegraphics[width=.49\textwidth]{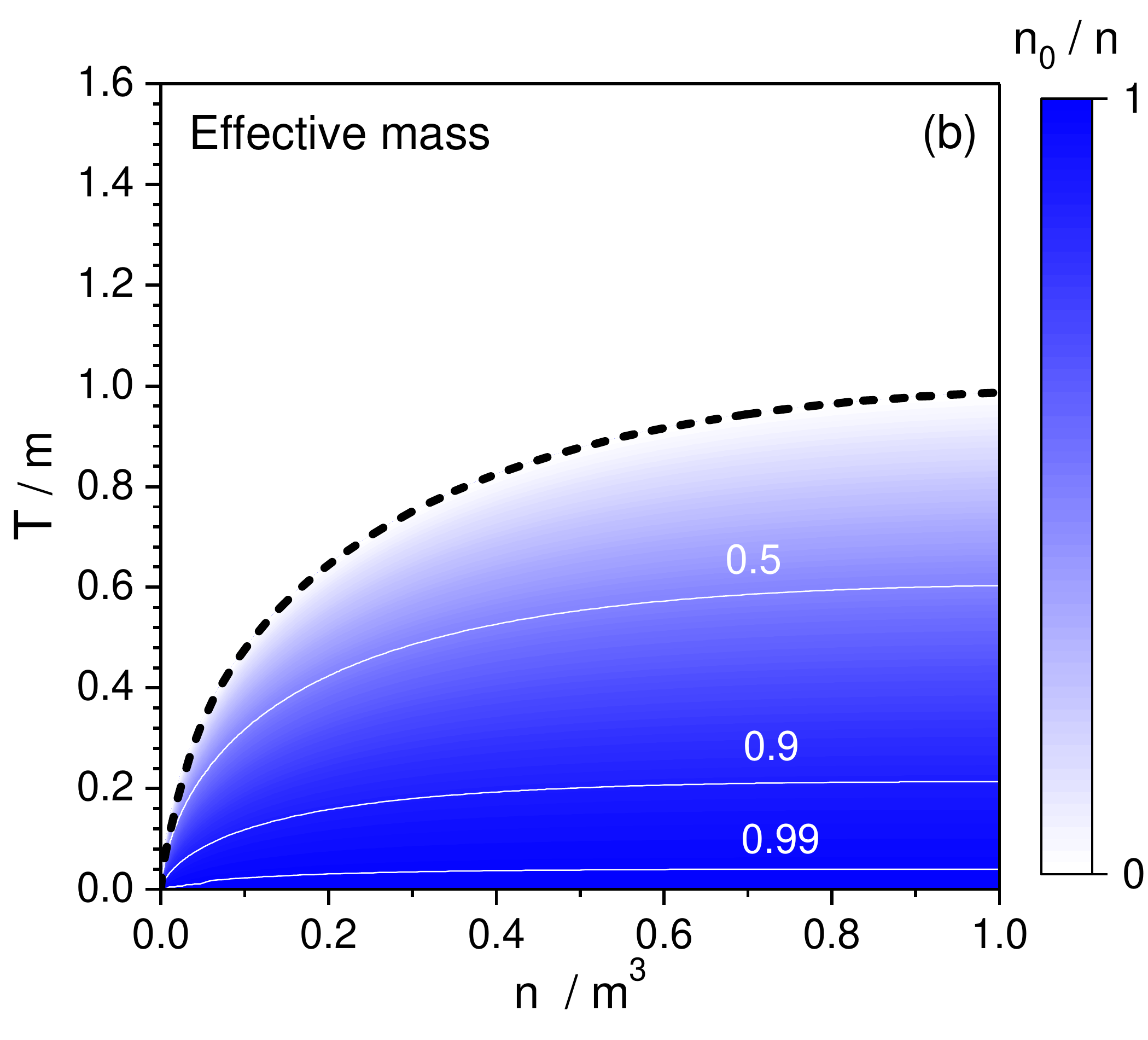}
\caption{\label{fig-cond-frac}
The BC fraction, $n_{0}/n$, as a function on $n$ and $T$ for the Id-BG and MF models  ($a$),  and for the EM model (b).  The line of the onset of BEC is shown by the dashed curve for each model. The result for the EV model is close to the Id-BG/MF result and, thus, is not presented.
}
\end{figure*}

The BC fraction,
\begin{equation} \label{Id-BC}
\frac{n_{0}}{n}=1-\frac{n_{\rm id}(T,\mu=m)}{n_{\rm id}(T_c,\mu=m)}~,
\end{equation}
lies between zero at the onset of the BEC at $T=T_c$ 
and unity at $T=0$. 
In the non-relativistic and ultra-relativistic cases this quantity takes respectively the following forms:
\eq{\label{cond-nr}
& \frac{n_0}{n}=1-\left(\frac{T}{T_c}\right)^{\frac{3}{2}},~~~~T_c/m\ll 1~; \\
& \frac{n_0}{n}=1-\left(\frac{T}{T_c}\right)^{3},~~~~T_c/m\gg 1~.\label{cond-rel}
}
The BC fraction $n_0/n$ in the Id-BG at $T<T_c$ is shown in Fig.~\ref{fig-cond-frac} (a). 

The BC does not produce 
an additional pressure in the Id-BG\footnote{This does not necessarily apply to systems of interacting bosons.}, thus the system pressure at the phase with BC equals $p_{\rm id}(T,\mu=m)$ at $T\le T_c$.
There is also no contribution from the BC to the entropy density,
\begin{equation}
s_{\rm id}(T, \mu=m)\equiv \left(\frac{\partial p_{\rm id}(T,\mu)}{\partial T}\right)_{\mu=m}~,
\end{equation}
whereas the
energy density $\varepsilon$  at $T\le T_c$ does get a contribution from the BC and reads
\begin{equation}
\varepsilon=\varepsilon_{\rm id}(T,\mu=m)+m\,n_0~.
\end{equation}
 At $T=0$ all particles of the Id-BG are in the BC state, thus 
\eq{\label{Id-BG-0}
n=n_0~,~~~p=0,~~~~s=0,~~~~\varepsilon=m~n_0.
}
 
\begin{figure*}
\includegraphics[width=.49\textwidth]{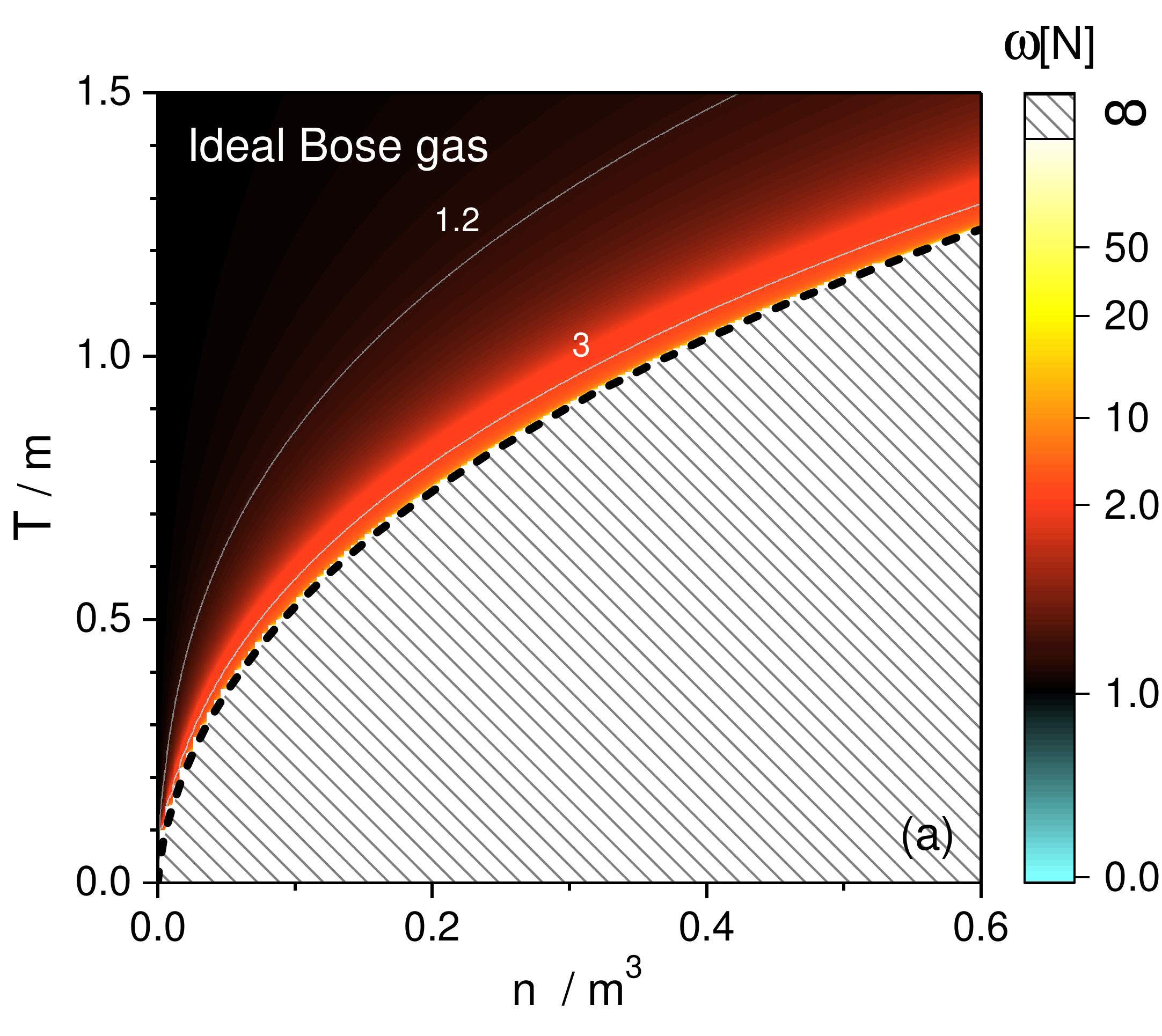}
\includegraphics[width=.49\textwidth]{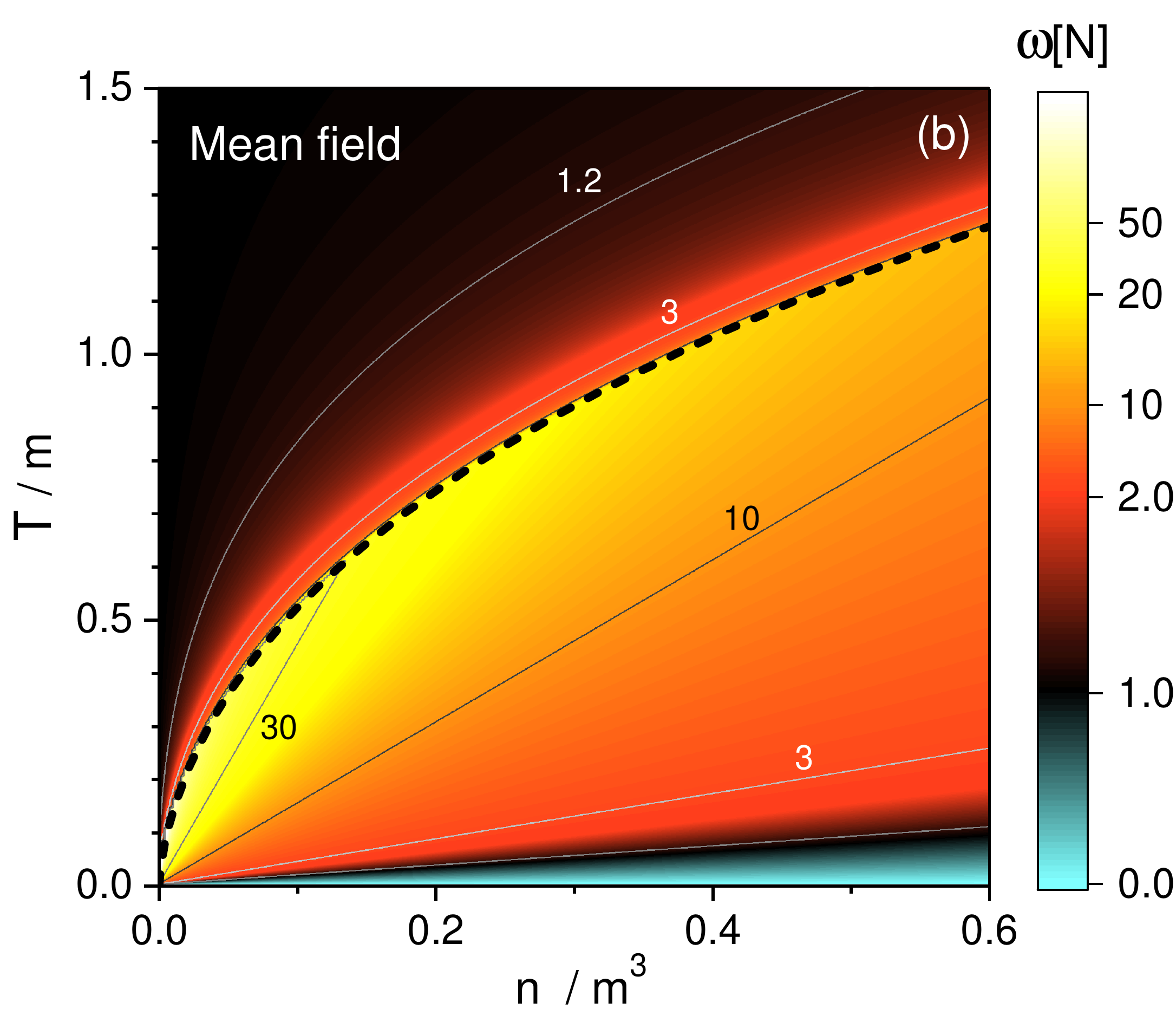}
\includegraphics[width=.49\textwidth]{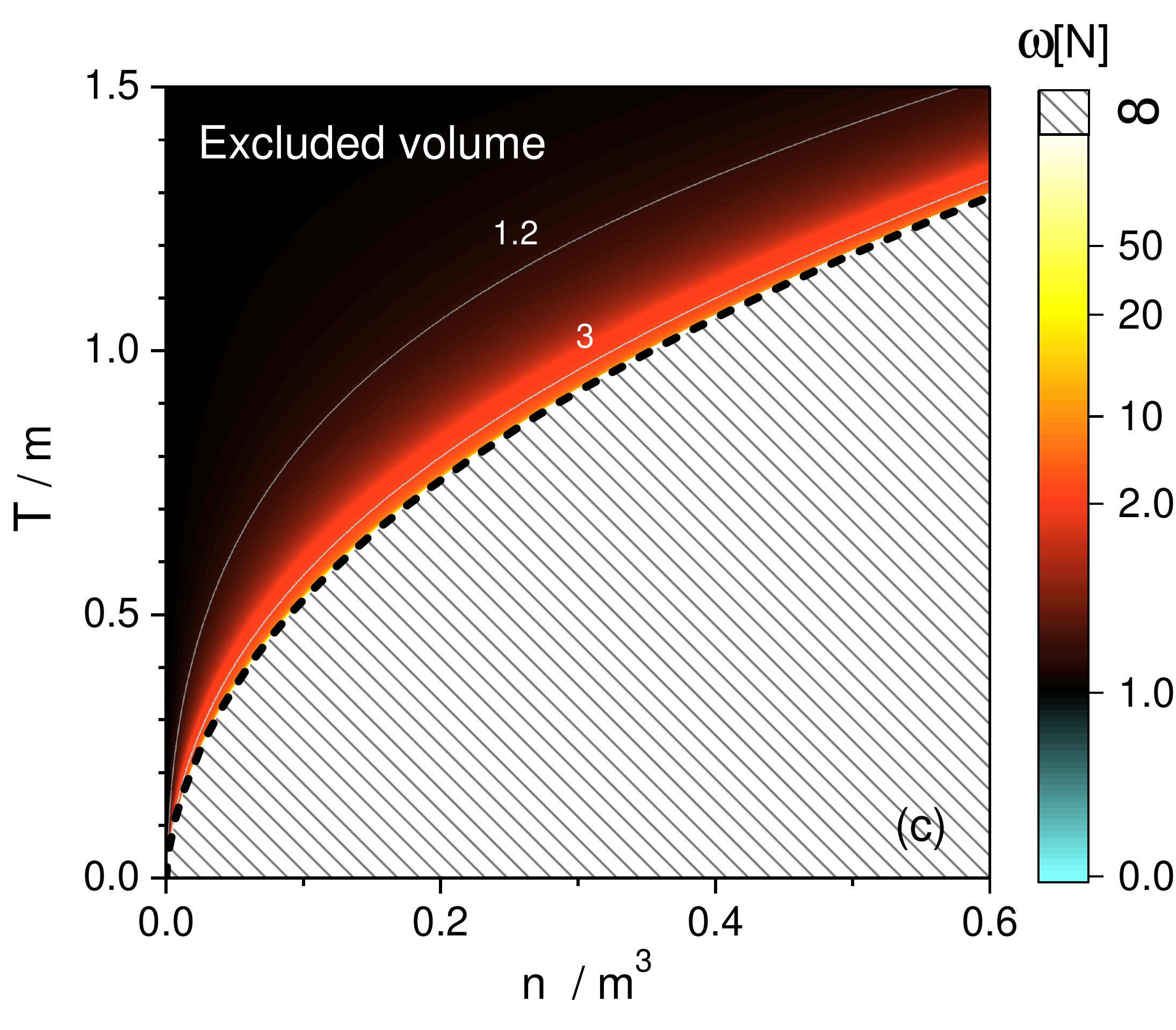}
\includegraphics[width=.49\textwidth]{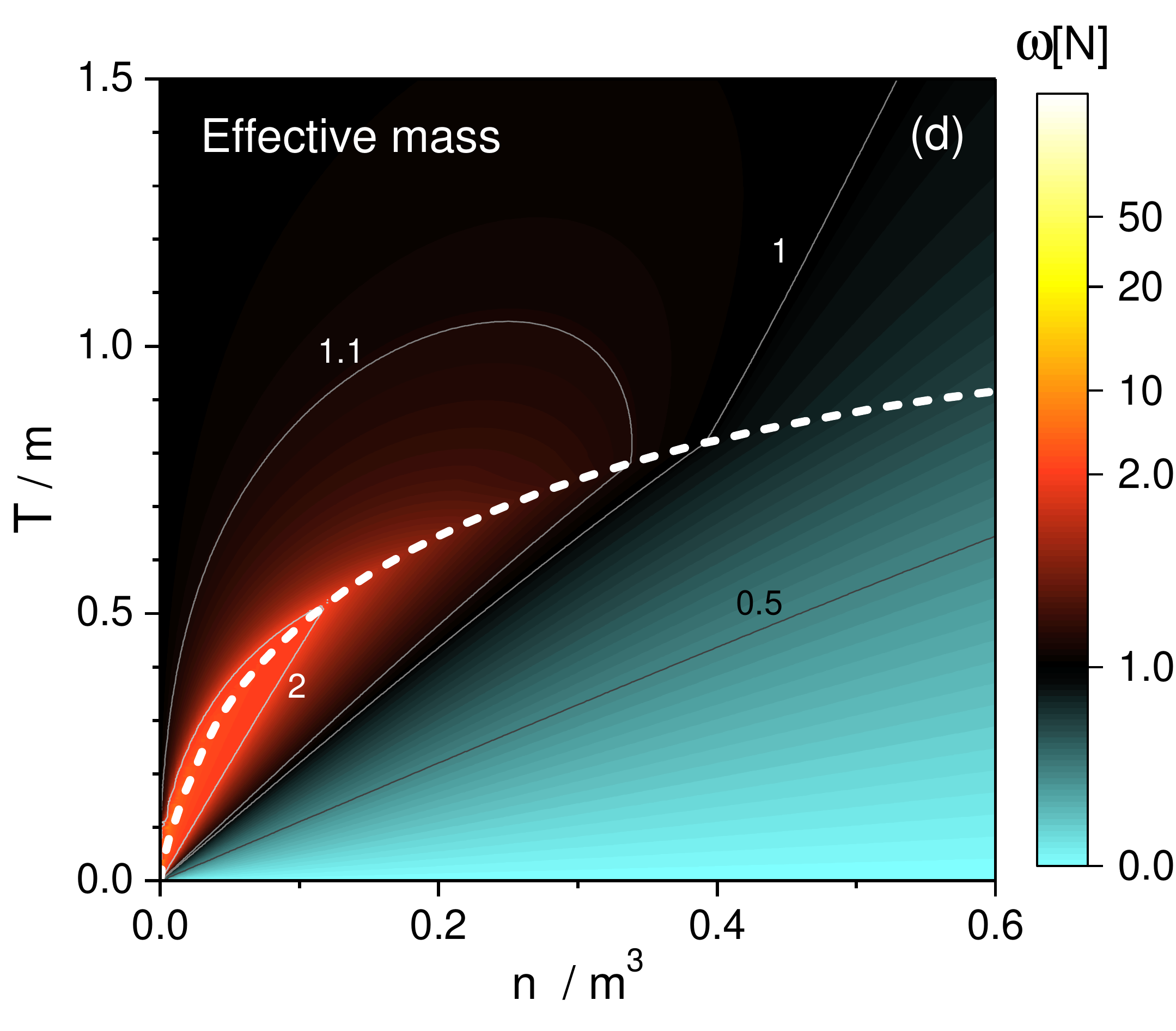}
\caption{\label{fig-w}
The scaled variance $\omega$ of particle number fluctuations as the function of density and temperature for the Id-BG (a),
the MF model (b), the EV model (c), and the EM model (d).  
The black colour corresponds to the $(n,T)$ region where $\omega$ is close to the Poisson limit,  $\omega=1$, i.e.  both the  Bose statistics and repulsive interaction effects are small. 
Regions with infinite value of  $\omega$ in pallets (a) and (c) are hatched. The line of the onset of BEC is shown by the dashed curve for each model.
}
\end{figure*}

A useful measure for the particle number fluctuations
in the thermodynamic limit
is the scaled variance:

\eq{\label{omega-N}
\omega=\lim_{N\rightarrow\infty} \frac{\mean{N^2}-\mean{N}^2}{\mean{N}}~,}
where $N$ is the total number of particles, and $\langle \ldots \rangle$ denotes the grand canonical ensemble  averaging\footnote{The finite size effects have been discussed in Ref.~\cite{Begun:2008hq}}.
The scaled variance of particle number fluctuations in the Id-BG  is \cite{Begun:2006gj}
\eq{\label{eq:wid}
\omega & =
\frac{T}{n}\left(\frac{\partial n}{\partial \mu}\right)_T 
=
\omega_{\rm id} (T, \mu)   
\nonumber \\
&=~1~ +~
\frac{\eta }{2\,\pi^2\,n_{\rm id}} \int_0^{\infty} dk k^2
[f_{\rm k}(T,\mu,m)]^2~.
}
The scaled variance $\omega_{\rm id} (T, \mu)$ in the Boltzmann approximation ($\eta=0$) equals to $\omega_{\rm}=1$, meaning that the particle number distribution
in the classical ideal gas is given by a Poisson distribution.
For the Id-BG ($\eta=1$) the scaled variance $\omega_{\rm id}$ is always above unity, and for the ideal Fermi gas it is always smaller than unity. $\omega_{\rm id}\rightarrow 1$ in the $T$-$\mu$ regions of the phase diagram where the effects of Bose and Fermi statistics are negligible, i.e. when $f_k\ll 1$.  
For the Id-BG the scaled variance (\ref{eq:wid}) diverges as one approaches the BEC-line:
\eq{\label{omega-id-BC}
\omega_{\rm id}(T_c,\mu=m)=\infty~.
}
Relation (\ref{omega-id-BC}) remains valid in the phase with the BC, i.e. for all $T< T_c$. The scaled variance $\omega$ for the Id-BG is presented as a function of $T$ and $n$ in Fig.~\ref{fig-w} (a).

\section{Models of repulsive interactions}
\label{sec-frameworks}

To address the problem of BEC in the presence of repulsive interactions we will consider three phenomenological models. 
As motivated in the Introduction,
we shall refer to the system of $\pi$-mesons in our consideration from this point on\footnote{The obtained results, however, have general validity.}. The $\pi$-mesons are bosons with spin equal to zero. There are three types of pions, $\pi^+$, $\pi^-$, and $\pi^0$, with $m_{\pi^{\pm}}\cong 140 $~ MeV and $m_{\pi^0}\cong 135$~MeV. 
In the relativistic systems the number of particles is not a conserved quantity. 
In the case of pions, the conserved quantity is the electric charge~(or, equivalently, isospin). 
The average value of the electric charge is regulated by the electric chemical potential $\mu_Q$ in the grand canonical ensemble. 
The chemical potentials of all three pion species are defined by $\mu_Q$ only: 
$\mu_{\pi^+}=\mu_Q$, $\mu_{\pi^-}=-\mu_Q$, and $\mu_{\pi^0}=0$. 
Nonzero values of $\mu_{\pi^+}$ (or $\mu_{\pi^-}$) in chemically equilibrated systems are only possible for non-zero net electric charge.
Chemical non-equilibrium is another possibility, where fast non-equilibrium processes  can produce overpopulation of all type of pions in comparison to the state of chemical equilibrium. 
Such a possibility has been suggested within ultra-relativistic collisions of hadrons and/or nuclei~(see, e.g., \cite{Begun:2006gj}).
In what follows we focus on the BEC of a single pion species.
Therefore, we fix a single pion type with the degeneracy factor $d=1$ 
and particle mass $m=m_{\pi}$, keeping the notation
$\mu$ for the chemical potential of that pion species.

In this section we present the formulations of the three models under consideration for phase diagram regions without BEC.
In the next section we generalize these models to describe the region with a non-zero BC component.

{\bf Mean field model.} The first model under consideration is the thermodynamic mean field (MF) model~(see e.g., \cite{Anchishkin:2014hfa} and references therein).
Within the MF model the pressure and particle number density at given $\mu$ and $T>T_c$ are given by the following equations:
\eq{\label{MF-p}
p(T,\mu)&=p_{\rm id}(T,\mu^{*})+\int\limits_{0}^{n}dn'\,n' \frac{dU(n')}{dn'}~,\\
\label{MF-n}
n
&=\left(\frac{\partial p}{\partial \mu}\right)_T = n_{\rm id}(T,\mu^{*})~,\\
\label{MF-mu}
\mu^* & = \mu-U(n)~.
}
Here $ U(n)$ is a density-dependent mean field and  $\mu^*$ is the effective chemical potential. The purely repulsive interactions correspond to $dU/dn> 0$.
The MF model with 
\eq{\label{an}
U(n)=an~, ~~~~ a>0~,
}
was applied to describe the BEC of interacting bosons in Ref.~\cite{poluektov2017simple}. More elaborate potentials with higher powers of $n$ can be of interest at high densities,
the Skyrme-like MF
model with $U(n)$ describing both repulsive and attractive interactions
was used to study the BEC of $\alpha$-particles
in Refs.~\cite{Satarov:2017jtu,Satarov:2018gfi,Satarov:2019vox}. 
We use the simplest version of $U(n)$ in the form of Eq.~(\ref{an}) in the following.

The scaled variance $\omega$~(\ref{omega-N}) of particle number fluctuations in the MF model 
can be presented as 
\eq{\label{MF-w}
\omega=
\frac{T}{n}\left(\frac{\partial n}{\partial \mu}\right)_T
=\frac{\omega_{\rm id} (T, \mu^*)}{1+\ddfrac{a\,n}{T}~\omega_{\rm id} (T, \mu^*)}~.
}
In the Boltzmann approximation $\omega_{\rm id}=1$. From Eq.~(\ref{MF-w}) it then follows
\eq{\label{omega-cl}
\omega~=~\left(1~+~\frac{a\,n}{T}\right)^{-1}~<1~,
}
and $\omega\rightarrow 1$ at $a\rightarrow 0$. Therefore, in the classical (Boltzmann) gas the MF repulsion leads to a suppression of particle number fluctuations. A stronger repulsion (larger $a$) leads to the stronger suppression. In the Id-BG, on the other hand, 
$\omega$ is always larger than 1 due to the effects of Bose statistics.

{\bf Excluded volume model}. The next approach is 
the excluded volume (EV) model which describes repulsive interactions of particles in terms of their eigenvolumes.
This approximation is usually used to model short-range repulsion similar to the hard-core repulsion in a classical gas of hard spheres. 
The EV model was generalized to include the effects of quantum statistics in Ref.~\cite{Vovchenko:2017cbu}.
It is 
defined by the following equations:
\eq{
\label{eq:pqvdw}
& p(T,\mu)  =  p_{\rm id}(T,\mu^*), \\
\label{eq:nqvdw} 
& n  = \left(\frac{\partial p}{\partial \mu}\right)_T = 
\frac{ n_{\rm id}(T, \mu^*)}{1+b\,n_{\rm id}(T,\mu^*)}~, \\
\label{muEV}
& \mu^*=\mu~-~b\,p_{\rm id}(T,\mu^*)~,
}
where $b>0$ is the classical eigenvolume parameter which regulates the strength of repulsion. As seen from Eq.~(\ref{eq:nqvdw}) the particle number density of the EV model satisfies an inequality $n<1/b$. 
In the classical gas of hard spheres the parameter $b$ is identified with the second virial coefficient and is expressed through hard-core radius $r$ as $b=16\pi r^3/3$.
Note, however, that
in the case of quantum hard-spheres
the second virial coefficient suggests a temperature dependent $b$~\cite{Vovchenko:2017drx}, which will be addressed in future works.

The EV model scaled variance is calculated as follows~\cite{Vovchenko:2015pya}:
\eq{\label{omega-EV}
\omega(T,\mu) = 
\frac{T}{n}\left(\frac{\partial n}{\partial \mu}\right)_T= 
(1-bn)^2\,\omega_{\rm id} (T, \mu^*).
}
The scaled variance in the EV model is suppressed by the factor $(1-bn)^2$ in comparison to $\omega_{\rm id}$.
Therefore, for classical ideal gas, where $\omega_{\rm id}=1$, one finds $\omega = (1-bn)^2<1$, i.e. the  EV repulsion effects lead to a suppression of the particle number fluctuations \cite{PhysRevC.76.024901}.

{\bf Effective mass model}. The third model that we consider is the effective mass (EM) model. 
A  formulation of the EM  model with $m^*=m^*(T)$ 
and $\mu=0$
was suggested in Ref.~\cite{PhysRevD.52.5206}
(see also Refs.~\cite{Begun:2010eh,Begun:2010md}). 
We extend the EM model to $m^*=m^*(T,\mu)$ by choosing a simple modification of the model that leads to a thermodynamically consistent description:
\eq{
p(T,\mu)~ & =~p_{\rm id}(T,\mu;m^*)~+~\frac{\left( m-m^*\right)^2}{2c}\, , \label{p-EM}\\
 n(T,\mu)
~& =~n_{\rm id}(T,\mu;m^*)\, , \label{n-EM}\\
m^*(T,\mu)~& =~ m\, + \, c~n^s_{\rm id}(T,\mu ; m^*)~,\label{m-EM}
}
where
\eq{
 n^s_{\rm id}(T,\mu;m^*) =\frac{d~m^*}{2 \pi^2}
 \int\limits_0^{\infty} 
 \frac{ k^2dk}{\sqrt{k^2+m^{*2}}}~ 
 f_{\rm k}(T,\mu; m^*)~ \label{ns}  
 }
is the scalar density of an ideal Bose gas and $c>0$ is a model parameter.

The requirement $c>0$ leads to $m^*> m$, which corresponds to the repulsive interactions. The numerical value of $c$  regulates a strength of the particle repulsion. 
The EM model considered here resembles 
the Walecka model~\cite{Walecka:1974qa,Serot:1984ey,Poberezhnyuk:2017yhx}
of nuclear matter.
However, the second term in the right hand side of 
Eq.~(\ref{p-EM}) is positive and describes the pion repulsion, whereas in the Walecka model the corresponding term is negative and it describes the attractive forces between nucleons.

The scaled variance of the particle number fluctuations is calculated as follows:
 \eq{\label{em-w}
\omega  &= \frac{T}{n}\left(\frac{\partial n}{\partial \mu}\right)_T \nonumber \\
& = \frac{T}{n} \left[ \frac{\partial n_{\rm id}(T,\mu;m^*)}{\partial \mu} + \frac{\partial n_{\rm id}(T,\mu;m^*)}{\partial m^*} \, \frac{\partial m^* (T,\mu)}{\partial \mu}\right]~.
}
The partial derivative $\partial m^* (T,\mu)/\partial \mu$ is evaluated by differentiating Eq.~\eqref{m-EM} with respect to $\mu$ and solving the resulting equation for $\partial m^* (T,\mu)/\partial \mu$:
\eq{
\frac{\partial m^* (T,\mu)}{\partial \mu} = \frac{c~\ddfrac{\partial }{\partial \mu}n^s_{\rm id}(T,\mu;m^*)}{1 ~-~ c~\ddfrac{\partial }{\partial m^*}n^s_{\rm id}(T,\mu;m^*)}~.
}
The expression for $\omega$ simplifies in the Boltzmann approximation $\eta=0$, i.e
when effects of quantum statistics are neglected, and by applying either non-relativistic~($T/m^* \ll 1$) or ultra-relativistic~($T/m^* \gg 1$) limits:
\eq{
\omega &=\left(1+c\,\frac{n}{T}\right)^{-1} , ~~~~~T/m^* \ll  1~,\\
\omega &=1, ~~~~~~~~~~~~~~~~~~~~T/m^* \gg  1~.}

{\bf Tuning the model parameters.} The positive model parameters, $a$, $b$, and $c$ for MF, EV, and EM models, respectively, regulate the strength of repulsive interactions in all the models.  
At $a=b=c=0$ all the three models are reduced to the Id-BG.  
We fix the numerical values of the above parameters through the following considerations. 
At $\mu=0$ one observes a suppression of the system's pressure in comparison to the Id-BG due to the repulsion effects. These suppression effects become larger with increasing  $T$, and the ratio $p(T,0)/p_{\rm id}(T,0)$ thus decreases with $T$ in all three descriptions. 
To keep only small deviations from the Id-BG results in the whole temperature region for $\mu=0$ we fix this ratio at $m_{\pi}=135$~MeV, $T=150$~MeV to a value
\eq{\label{ratio}
\frac{p(T,\mu=0)}
{p_{\rm id}(T,\mu=0)}~\cong~0.98~< 1~,
}
the same for all three models.
Therefore, the difference between all three models is almost negligible at $\mu=0$, and their deviations from the Id-BG pressure are indeed very small at all physically reasonable $T$ values\footnote{
The QCD chiral crossover transition pseudocritical temperature at $\mu=0$ is
$T_{\rm pc} \cong 155~{\rm MeV}$~\cite{Bazavov:2018mes,Borsanyi:2020fev}.}.
The requirement (\ref{ratio}) fixes the model parameters to
\eq{
 \text{MF model:} \quad & \quad a=0.15~m_{\pi}^{-2}~,\\
 \text{EV model:} \quad & \quad b=0.145~m_\pi^{-3}~, \\ 
 \text{EM model:} \quad & \quad c=2.21~m_{\pi}^{-2}. 
 }
If $b$ is to be interpreted as the 
excluded volume parameter in the system of classical  spheres, one finds 
$r\cong 0.3$~fm for the pion ``hard-core" radius. 
This is consistent with the values considered in Refs.~\cite{Andronic:2012ut,Vovchenko:2014pka,Anchishkin:2018fab}.

One comment is appropriate here. An evident intuitive expectation is that the system's pressure should increase when the inter-particle repulsion is switched on. Equation (\ref{ratio}) demonstrates the opposite behavior. 
This counter-intuitive result comes due to a decrease of particle number density at any fixed $T$ and $\mu$ values as a consequence of the repulsive interactions. This suppression of the particle number density $n$ leads to the lower values of the  pressure when compared with the corresponding values in the system of non-interacting particles at the same $T$ and $\mu$ (see Ref.~\cite{Yen:1997rv}).
The pressure in models with repulsive  interaction becomes indeed higher then that of the Id-BG  if they will be compared at fixed $T$ and $n$ values. Therefore, the ratio  
\eq{\label{ratio-1}
\frac{p(T, n)}
{p_{\rm id}(T, n)}~>1~
}
is larger than 1 for all three models. For example, at $T=110$~MeV and $n=0.06~{\rm fm}^{-3}$ the numerical values of the ratio (\ref{ratio-1}) equal approximately  to $1.02$, $1.02$, and $1.15$ for MF, EV, and EM models, respectively.

\section{BEC analysis}
\label{sec-results}
{\bf MF model.}
In the MF model~[Eqs.~(\ref{MF-p}-\ref{an})] the BEC-line corresponds to the effective chemical potential being equal to the particle's mass, $\mu^*=m$.
Therefore, as follows from Eq.~(\ref{MF-n}) the density, $n$, and temperature, $T_c$, at the BEC-line are connected by the same equation as for the Id-BG case:
\eq{\label{eq:MF-CP}
n =n_{\rm id}(T_c,\mu^*=m),
}
Thus, the BEC-line in the MF model coincides with that in the Id-BG in the $(n,T)$ plane, and it is described by the same analytic relations as in Eqs.~(\ref{nrel}) and (\ref{rel}). The MF model BEC-line is shown in Fig.~\ref{cond-lines} (a) by the dashed red ine.
The BC fraction also keeps its Id-BG forms (\ref{cond-nr}) and (\ref{cond-rel}). This is depicted in Fig.~\ref{fig-cond-frac} (a).
However, as follows from Eq.~(\ref{MF-mu}), the chemical potential at $T\le T_c$ behaves as follows:
\eq{\mu ~=m~+~U(n)\label{MF-n-of-mu}}
and does not keep a constant value along the-BEC line, 
in contrast to $\mu= m$ constant value in the Id-BG. 
Namely, the $\mu = \mu_c$ value increases along the BEC-line with increasing $T_c$ as shown in Fig.~\ref{cond-lines} (b). 
This is because the critical density $n=n_c$~(and thus the $U(n_c)$ contribution to $\mu_c$ in Eq.~\eqref{MF-n-of-mu}) increases with $T_c$ as seen from Fig.~\ref{cond-lines} (a).  

At $T\le T_c$ Eq.~(\ref{MF-n}) is modified to account for the non-zero contribution from the BC:
\eq{
n =n_{\rm id}(T,\mu^*=m)+n_0~,~~~~~~T\leq T_c~.
}
This can be used to calculate the BC density $n_0$ as a function of $n$ and $T$. The BC density satisfies the same equation (\ref{Id-BC}) as in the Id-BG, shown in Fig.~\ref{fig-cond-frac} (a). 
It leads to $0\le n_0\le n$ with $n_0=0$ at $T=T_c$ and $n_0=n$ at $T=0$.

The scaled variance of particle number fluctuations is given by Eq.~\eqref{MF-w}.
Approaching the BEC-line ($\mu^*= m$) one has $\omega_{\rm id}(T,\mu^*)\rightarrow  \infty$ in Eq.~\eqref{MF-w} and thus
\eq{\label{omega-MF-BEC}
\omega =\frac{T}{n}\left(\frac{dU}{dn}\right)^{-1}= \frac{T}{an}~,\qquad T\leq T_c 
}
Note that Eq.~(\ref{omega-MF-BEC}) remains valid at all $T\le T_c$.

The scaled variance for 
the MF model is presented in Fig.~\ref{fig-w} (b). 
In contrast to the Id-BG case, in the MF model $\omega$ is finite  both near the BEC-line and inside a phase with the BC, $n_0>0$,
changing continuously as one goes across the BEC-line.
Equation (\ref{MF-w}) illustrates also the requirement of thermodynamic stability of the MF model for bosons, especially on the BEC-line. To fulfil an evident requirement $\omega >0$, which follows from the definition (\ref{omega-N}), one needs $dU/dn>0$. This last inequality is valid for repulsive interactions.   In particular, this is valid for $U(n)$ given by Eq.~(\ref{an}) with $a>0$. Purely attractive interaction with $a<0$ in Eq.~(\ref{an}) would lead to thermodynamic instability.

{\bf EV model.}
The temperature $T_c$ on the BEC-line is obtained by substituting $\mu^*=m$ in Eq.~(\ref{eq:nqvdw}) and solving that equation with respect to $T$.
In contrast to the MF model,  
the EV model $T_c(n)$ dependence does not coincide with the Id-BG result. 

At $T= T_c$ Eqs.~(\ref{eq:nqvdw}) 
and (\ref{muEV}) are transformed to
\eq{
n&=\frac{n_{\rm id}(T_c,\mu^*=m)}{1+b\,n_{\rm id}(T_c,\mu^*=m)}~, \label{nEV-1}\\
\mu&=m+bp_{\rm id}(T_c,\mu^*=m)~. \label{muEV-1}
}
The solutions of Eqs.~(\ref{nEV-1}) and (\ref{muEV-1}) are shown in Figs. \ref{cond-lines} (a) and (b), respectively. A novel feature of the EV model is the modification of the BEC-line in the $(n,T)$ plane with $T_c^{\rm EV}(n)$ being larger than $T_c^{\rm MF}(n)=T_c^{\rm id}(n)$ BEC-line of the MF and Id-BG models. 
This modification of the BEC-line with respect to the Id-BG result, shown in Fig.~\ref{cond-lines} (a) is consistent, 
with many other results reported for Bose gases with repulsion~\cite{PhysRevLett.83.1703,Baym_2000, Holzmann_1999,Holzmann_2001,Baym_1999,PhysRevLett.83.3770,PhysRev.91.1291,PhysRev.91.1301}.

The scaled variance $\omega$ is given at $T \geq T_c$ by Eq.~\eqref{omega-EV}.
As one approaches the BEC-line~($\mu^* \to m$), the Bose effects start to dominate $\omega$ since $\omega_{\rm id}(T,\mu^* \to m)\rightarrow \infty$, implying that particle number fluctuations start to diverge.
The scaled variance remains divergent at all $T\le T_c$ in the EV model.
This behavior is similar to the Id-BG but differs from the MF model where fluctuations remain finite everywhere.
The behavior of $\omega$ as a function of temperature and density in the EV model is presented in Fig.~\ref{fig-w}~(c).

\begin{figure}
\includegraphics[width=.49\textwidth]{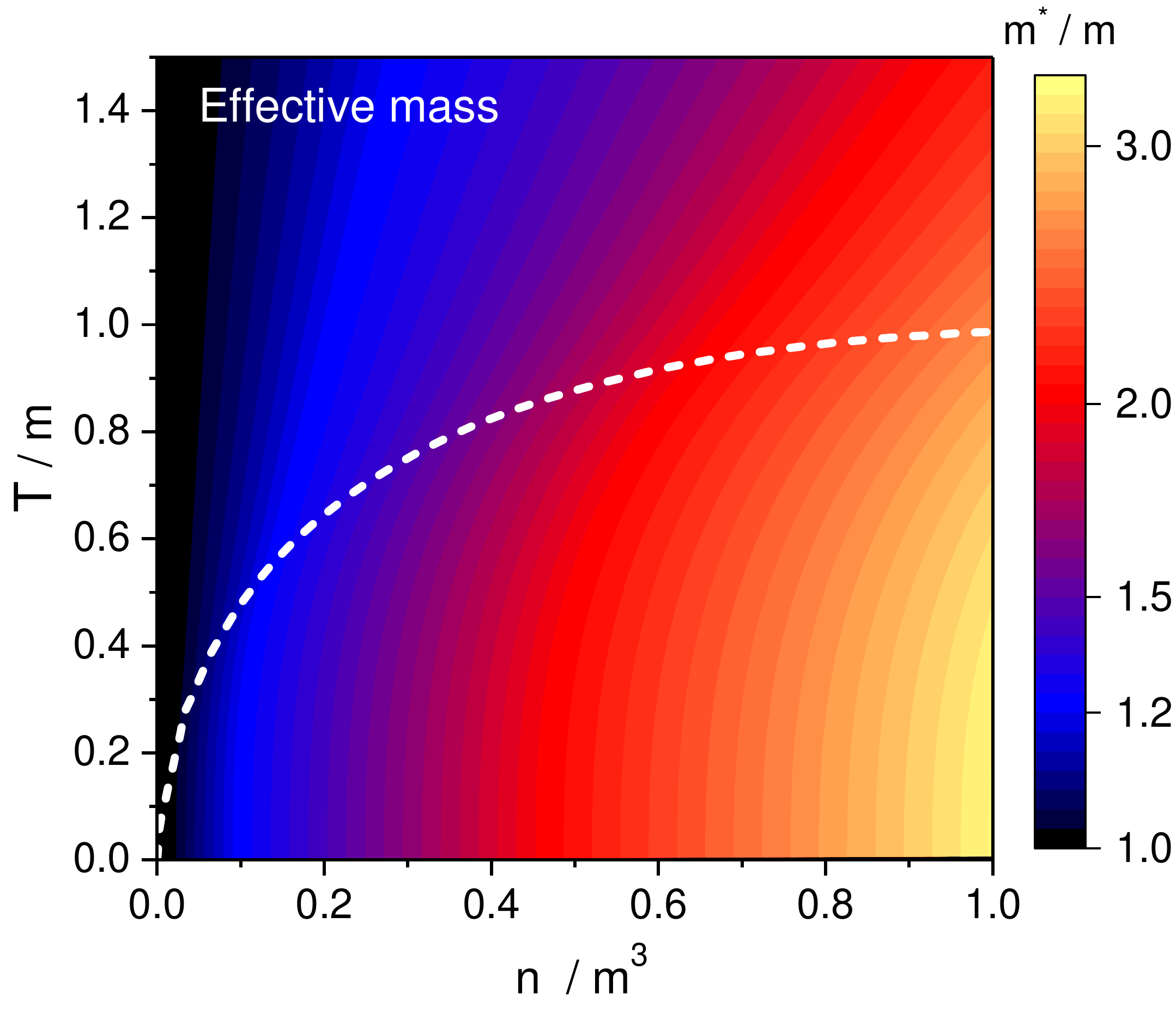}
\caption{\label{fig-em}
Effective mass $m^*/m$ as a function of density and temperature for the EM model.  
The black colour corresponds to the $m^*=m$ region where both the Bose statistics and repulsive interaction effects are negligible.  The line of the onset of BEC is shown by the dashed curve.
}
\end{figure} 

{\bf EM model}.
The phase with the BC corresponds in the EM model to a condition $\mu = m^*$. 
The BEC-line is thus defined by the following equation
\eq{\label{m*}
m^*(T_c)=m+ c\,n_s(T_c,\mu=m^*;m^*)~.
}
The resulting BEC-line $T_c=T_c(\mu)$ is shown in Fig.~\ref{cond-lines} (b) by a dotted line. 
With $n$ calculated from 
\eq{\label{nc-EM}
n=n_{\rm id}(T_c,\mu=m^*;m^*)
}
the BEC-line is shown in Fig.~\ref{cond-lines} (a) as a function of particle number density, $T_c=T_c(n)$ . 
A distinct new feature of the EM model is a decrease of the BEC-line $T_c(n)$ as compared to the Id-BG. Therefore, the three considered models of particle repulsion show three qualitatively different possibilities for the changes of the BEC-line in comparison to the Id-BG behavior, namely
\eq{\label{Tc}
T_c^{\rm EV}(n)~>~T_c^{\rm MF}(n)~=~T_c^{\rm id}(n)~>~T_c^{\rm EM}(n)~.
}
Another interesting feature of the  EM model is an absence 
of the BEC at large temperatures, i.e. 
Eq.~\eqref{m*} has no solutions at $T > T_c^{\rm max}$, meaning that
the BEC is only possible at $T< T_c^{\rm max}$. 
The $T_c^{\rm max}$ value is obtained by numerically analyzing Eq.~(\ref{m*}) and it is marked  by the stars in Fig.~\ref{cond-lines}.  
Explicit expressions for $T_c^{\rm max}$ can be obtained if the non-relativistic or ultra-relativistic approximations are applied:
\eq{\label{Tmnr}
&T_c^{\rm max} \cong ~
\frac{2\pi m}{3}\left(\frac{2}{m^2c\,\zeta(3/2)}\right)^{2/3}~,        ~~~m^2c \gg 1~, \\
&T_c^{\rm max} \cong~~ m\left(\frac{12}{m^2c}\right)^{1/2},   ~~~~~~~~~~~~~\quad m^2c \ll  1~. \label{Tmr}
}

The expressions~\eqref{p-EM}-\eqref{m-EM} for the EM model equation of state are modified in a presence of the non-zero BC density $n_0 > 0$ at $T< T_c$. Namely
\eq{
p(T,\mu)~ & =~p_{\rm id}(T,\mu;m^* = \mu)~+~\frac{\left( \mu-m\right)^2}{2c}\, , \label{p-EM-mod}\\
n(T,\mu)
~& =~n_{\rm id}(T,\mu;m^*=\mu) + n_0(T,\mu), \label{n-EM-mod}\\
n_0(T,\mu)~& =~ \frac{\mu - m}{c} - n^s_{\rm id}(T,\mu; m^* = \mu).
\label{m-EM-mod}
}

The BC density $n_0=n_0(n,T)$ in the EM model is presented in Fig.~\ref{fig-cond-frac} (b) whereas the behavior of the scaled effective mass $m^*/m$ presented in Fig.~\ref{fig-em}, both above and below $T_c$.

The scaled variance of particle number fluctuations 
for $T\geq T_c$ is given by Eq.~\eqref{em-w} while for $T<T_c$ it is evaluated as follows:
\eq{\label{em-w-mod}
\omega  &= \frac{T}{n}\left(\frac{\partial n}{\partial \mu}\right)_T 
= \frac{T}{n} ~\Big[~ \frac{\partial n_{\rm id}(T,\mu;m^*=\mu)}{\partial \mu} \\
& -~ \frac{\partial n^s_{\rm id}(T,\mu;m^*=\mu)}{\partial \mu} 
 + \frac{1}{c}~\Big]~.\nonumber
}
The function 
$\omega=\omega (n,T)$ in the EM model is presented in
Fig.~\ref{fig-w} (c).
We used Eq.~\eqref{em-w} for calculations at $T\geq T_c$ and Eq.~\eqref{em-w-mod} for calculations at $T < T_c$.
The scaled variance $\omega$ remains finite at all $(n,T)$ values. 
Furthermore, the behavior of $\omega$ is found to be continuous as one crosses the BEC-line.
This EM model result is similar to the MF model, but differs the Id-BG and EV results where $\omega$ is divergent for all $T\le T_c$.

The scaled variance $\omega$ is close to unity at fixed temperature $T>T_c$ and small density, $n \approx 0$, in all considered models of particle repulsion.
This is a region of the phase diagram where both the interaction and Bose statistics effects become negligible and the ideal gas Boltzmann approximation can be applied. 
However, close to the BEC-line and inside the phase with the non-zero BC $n_0>0$ the models differ qualitatively. 
In contrast to the Id-BG and EV models, one finds within the MF and EM models the finite values of $\omega$ at all temperatures and densities including those at $T\le T_c$.
These two models lead also to $\omega\rightarrow 0$ at $T\rightarrow 0$ as seen from Figs.~\ref{fig-w} (b) and (c).

In the limit $T \to 0$ the entropy approaches zero, $s \to 0$,
in all considered models. 
This is in agreement with the third law of thermodynamics. 
Also, all particles are located in the condensate at zero temperature, i.e. $n \to n_0$ as $T \to 0$, as is the case in the Id-BG. 
However, the pressure and energy density does depend on the specific model used. 
For instance, the
MF model yields the following in the zero temperature limit;
\eq{\label{zerotMF}
p~=~a~\frac{n_0^2}{2},~~~~\varepsilon=m\,n_0+a~\frac{n_0^2}{2}~.
}
The EM model gives similar expressions:
\eq{\label{zerotEM}
p=c\frac{n_0^2}{2},\quad \varepsilon=m\,n_0+c\,\frac{n_0^2}{2}.~
}
Both Eq.~\eqref{zerotMF} and \eqref{zerotEM} differ from the Id-BG result~(\ref{Id-BG-0}).

\section{Summary}
\label{sec-concl}

The phenomenology of the Bose-Einstein condensation (BEC) in equilibrium systems with repulsive interactions has been studied in three different models. The results from different models are  compared with each other as well as with the ideal Bose gas (Id-BG) baseline.
The mean field (MF), excluded volume (EV), and effective mass (EM) models have been considered  to describe the particle repulsion.
The model parameters are fixed to have very similar results at $\mu=0$ for all the considered temperatures. 
In this region of the phase diagram, far away from the BEC onset, the three considered models demonstrate {\it universal} features of the repulsive interactions, with only minor~($\sim2\%$) deviations of the system pressure from that in the Id-BG. However, the model results differ significantly in their {\it peculiar} behavior on the BEC-line as well as inside the phase with a Bose condensate.   

First,
deviations of the BEC-line $T_c=T_c(n)$ from the Id-BG baseline are qualitatively different in all three considered models: 
$T_c^{\rm EV}(n)>T_c^{\rm id}(n)$ for the EV model,
$T_c^{\rm EM}(n)<T_c^{\rm id}(n)$ for the EM model,
and $T_c^{\rm MF}(n)~=~T_c^{\rm id}(n)$ for the MF model.

Second, essential qualitative differences are observed for the behavior of the scaled variance $\omega$ of  particle number fluctuations. 
$\omega$ is divergent at $T\le T_c$ in the EV model, which is similar to the Id-BG behavior. 
On the other hand, the values of $\omega$ remain finite and continuous at all densities $n$ and temperatures $T$ within the MF and EM models. 
This fact provides an
opportunity to distinguish the features of particle interactions
experimentally, by the measurements of particle number fluctuations. 

Third, the EM model exhibits a distinctive feature:
existence of the maximal temperature $T^{\rm max}_c$ above which the BEC does not occur. 
Interestingly, a similar behavior is seen in lattice QCD simulations at finite isospin, with $T^{\rm max}_c \approx 160~\text{MeV} \approx 1.2~m_\pi$~\cite{Brandt:2017oyy}.
There, the disappearance of the pion BEC is usually attributed to a transition to partonic degrees of freedom -- a mechanism which the EM model studied here does not possess.
Nevertheless, the EM model could be useful for parameterizing the QCD equation of state at finite isospin density for temperatures $T \lesssim 160$~MeV.

\section*{Acknowledgments}
We are thankful to D.V. Anchishkin, I.N. Mishustin, and L.M. Satarov for fruitful discussions. This work is partially supported 
by the Target Program of Fundamental Research of the Department of Physics and Astronomy of the National Academy of Sciences of Ukraine
(N~0120U100857).
O.St. acknowledges the financial support by the scientific program “Astronomy and space physics” (Project N. BF19-023-01) of Taras Shevchenko National University of Kyiv.
V.V. was supported by the
Feodor Lynen program of the Alexander von Humboldt
foundation and by the U.S. Department of Energy, 
Office of Science, Office of Nuclear Physics, under contract number DE-AC02-05CH11231.
H.St. acknowledges the support through the Judah M. Eisenberg Laureatus Chair by Goethe University  and the Walter Greiner Gesellschaft, Frankfurt.

\bibliography{references.bib}
\end{document}